\documentclass[pra,twocolumn,showpacs,superscriptaddress,floatfix]{revtex4}
\usepackage[dvipdfm]{graphicx}
\usepackage{subfigure}
\usepackage{epsfig}
\usepackage{float}
\usepackage{ifpdf}
\ifpdf
	\usepackage[pdftex]{graphicx}
	\usepackage{epstopdf}
\else
	\usepackage[dvipdfm]{graphicx}
\fi
\usepackage{amsmath}	

\newcommand{\ket}[1]{\left | #1 \right \rangle}
\def\k(#1){|#1\rangle}
\newcommand{\bra}[1]{\left \langle #1 \right |}
\newcommand{\proj}[1]{\ket{#1} \bra{#1}}

\newcommand{\amp}[2]{\left \langle #1 | #2 \right \rangle}

\newcommand{\beq}{\begin{equation}}
\newcommand{\eeq}{\end{equation}}
\newcommand{\beqa}{\begin{eqnarray}}
\newcommand{\eeqa}{\end{eqnarray}}
\newcommand{\beqan}{\begin{eqnarray*}}
\newcommand{\eeqan}{\end{eqnarray*}}

\newcommand{\affA}{%
\affiliation{
 National Institute of Information and Communications Technology,
 4-2-1 Nukui-kita, Koganei, Tokyo 184-8795, Japan}
     }
\newcommand{\affB}{%
\affiliation{
 Sophia University,
 7-1 Kioicho, Chiyoda-ku, Tokyo 102-8554, Japan}
}
\newcommand{\affC}{%
\affiliation{
 Department of Information Engineering, University of Padua,
 Via Gradenigo 6/B, 35131, Padova, Italy}
}

\begin{document}

\title{Displacement receiver for phase-shift-keyed coherent states}
\date{\today}

\author{Shuro Izumi }
\affA%
\affB%
\author{Masahiro Takeoka}%
\affA%
\author{Mikio Fujiwara}%
\affA%
\author{Nicola Dalla Pozza}%
\affC%
\author{Antonio Assalini}%
\affC%
\author{Kazuhiro Ema}%
\affB%
\author{Masahide Sasaki}%
\affA%

\pacs{03.67.Hk, 03.67.-a}

\begin{abstract}
Quantum receiver is an important tool for overcoming
the standard quantum limit (SQL) of discrimination errors
in optical communication.
We theoretically study the quantum receivers
for discriminating ternary and quaternary
phase shift keyed coherent states
in terms of average error rate and mutual information.
Our receiver consists of on/off-type photon detectors and
displacement operations w/o electrical feedforward
operations.
We show that for the ternary signals, the receiver shows
a reasonable gain from the SQL even without feedforward. This scheme   
is realizable with the currently available technology. For the quaternary signals feedforward operation is crucial to overcome the SQL with imperfect devices. We also analytically examine the asymptotic limit of
the performance of the proposed receiver with respect to the number of
feedforward steps.
\end{abstract}

\maketitle

\section{Introduction}
Coherent states are not orthogonal to each other and then they can not be discriminated without errors. However, coherent states have special importance for communications since they are the best signal carriers. In fact, in most practical optical channels where energy loss is linear, they can propagate intact in pure states.
This characteristic is clearly emphasized in the theory on the ultimate capacity of a lossy bosonic channel
\cite{Giovannetti04}, which proves that the optimal encoding scheme, to attain the ultimate capacity under power constraint, has to employ a sequence of coherent-state pulses to code the information messages. Hence, the use of non-classical states at the transmitter
does not increase the channel capacity.
Quantum effects are required at the receiver
since optimal decoding generally calls for
entangling operations over a sequence of coherent states
\cite{Sasaki97,Sasaki98_superadd,Sasaki98_realization,Buck02,
Guha2011_PRL}.
The concept of `quantum collective decoding'
was first demonstrated in \cite{Fujiwara2003_PRL_ChannelCoding} where
polarization-location coding in a single photon was adopted.
Recently codeword demodulation for coherent states
without entangling operation was also
demonstrated \cite{Chen2012_NatPh} based on
conditional pulse nulling \cite{Guha2011_JMO}.
However, there are still technical challenges to realize
a quantum collective decoder for coherent states.

An important step towards this goal consists in realizing
a quantum optimal receiver that is not collective
but that can discriminate each single coherent state with minimum error probability.
The discrimination error in conventional receivers,
homodyne and heterodyne receivers, is bounded by the shot noise limit,
which is often referred to as the
{\it standard quantum limit} (SQL) in coherent optical communication.
On the other hand, Helstrom provided a theory to find the ultimate
lower bound to the error probability \cite{Helstrom_book76_QDET}. The Helstrom bound results to be exponentially lower
than the SQL and thus many efforts have been devoted to explore
how to design practical receivers able to approach such a limit.

For binary signals, it was shown that the SQL can not be outperformed
by Gaussian operations (up to the second order
optical nonlinear processes) with any classical conditional dynamics
\cite{TakeokaSasaki2008_DisplacementRec_GaussianLimit},
while the Helstrom bound is attainable if higher order
nonlinearities are freely available \cite{Sasaki96_Unitary_Control}.
One of the currently feasible ways to realize nonlinearities relies on
using highly efficient photon counters.
Dolinar proposed an optimal receiver composed by a displacement operation, a photon counter and feedback \cite{Dolinar73}; its performance was
demonstrated for the discrimination of on-off keying signals
\cite{CookMartinGeremia2007_Nature}. Furthermore, sub-optimal receivers without feedback consisting only of photon counting and optical displacement have been also proposed and experimentally demonstrated
\cite{Kennedy73,TakeokaSasaki2008_DisplacementRec_GaussianLimit,
Wittmann2008_PRL_BPSK,Tsujino2010_OX_OnOff}. The advantage of adopting such a simpler setup is that it easily allows to use a highly efficient photon detector such as a transition-edge sensor (TES) \cite{Lita2008_OX_TES_DE95,Fukuda2009_Metrologia_TES}. In
\cite{Tsujino2011_Q_Receiver_BPSK} it was demonstrated that without correcting any imperfection, with such a simpler architecture it is possible to outperform the SQL (the homodyne limit).

Respect to the binary case, much less attention has been paid to
the discrimination of $M$-ary signals with $M>2$.
Bondurant extended the Dolinar receiver to quaternary phase shift keying (4PSK) signals, and he proposed a sub-optimal receiver
consisting of continuous photon counting and infinitely fast
electrical feedback \cite{Bondurant93}. Recently some simpler schemes
have been proposed and experimentally tested. In \cite{Mueller2012_NJP} it was considered an hybrid scheme composed of homodyne and optimized displacement receivers with feedforward. In \cite{Becerra_NIST_2011_MPSK_emulation_experiment} displacement receivers and feedforward were employed and it was numerically showed that the proposed scheme is applicable for general $M$-ary PSK signals. Such a solution is particularly attractive since good performances can be achieved with a few number of feedforward stages and with moderated detection efficiency requirements. In the reported experiments, however, the dynamical feedforward was not performed (but `emulated'
\cite{Becerra_NIST_2011_MPSK_emulation_experiment}),
which indicates that there are still technical difficulties
to realize realtime feedforward in quantum receivers. In addition, in  \cite{Becerra_NIST_2011_MPSK_emulation_experiment}
it was left as a future task the discussion of the scalability of the proposed scheme with the number of feedforward steps $N$.

In this paper, we theoretically investigate
the displacement-based receiver for ternary and quaternary PSK signals, i.e., 3PSK and 4PSK.
Compared to previous works \cite{Bondurant93,Mueller2012_NJP,
Becerra_NIST_2011_MPSK_emulation_experiment}, our contribution includes
the following additional aspects.
First, we show that even with a simple setup
without any feedforward or feedback,
it is possible to overcome the SQL (the heterodyne limit).
Although the novel receiver requires relatively high detection efficiencies for photon counting, its implementation is feasible with state-of-art photon detectors, e.g. TES reported in
\cite{Lita2008_OX_TES_DE95,Fukuda2009_Metrologia_TES}.
Second, we provide analytical expressions for
the error rate performance of
the displacement receiver with feedforward, which structure is basically
similar to the setup given in \cite{Becerra_NIST_2011_MPSK_emulation_experiment}.
We show that the adoption of feedforward operations drastically
improve the error rate performance, and consequently tolerate the requirement for photon detectors,
in agreement with the results
in \cite{Becerra_NIST_2011_MPSK_emulation_experiment}.
In addition the obtained analytical formula allows one to clarify
the scalability of the performance in the limit of large $N$.
We also compare the performance with the Bondurant receiver \cite{Bondurant93} and the Helstrom limit.
Finally, we present an analysis based
on the mutual information of the system
including a comparison with the unambiguous state discrimination method.

This paper is organized as follows.
In Sect.~\ref{Sect:2}, we discuss the displacement receiver
without feedforward.
The performance of the receivers with feedforward are
analyzed in Sect.~\ref{Sect:3}.
Sect.~\ref{Sect:4} is devoted to the mutual information analysis
and the paper is concluded in Sect.~\ref{Sect:5}.

%
\section{Displacement receiver without feedforward}\label{Sect:2}
%

In this section we shall propose and describe the structure of
two receivers, which do not include any feedforward operation,
for the 3PSK and 4PSK signals.

The $M$--ary PSK coherent states $\ket{\alpha_m}$, $m=0,1,\ldots,M-1$,
are defined as \beq \ket{\alpha_m}=\ket{\alpha\, u^m},\quad u=e^{2\pi i / M}\;,\eeq
where, without loss of generality,
$\alpha$ is chosen to be a real number.
Throughout this paper, we assume that the a-priori probabilities of the signals are all the same, i.e. equal to $1/M$.
The states can be generated as
\beq\label{V_def}
\ket{\alpha_m}=\hat V^m \ket{\alpha_0},\quad
\hat V=\exp\left( \frac{2\pi i}{M} \hat n \right)\;,
\eeq
where
$\hat{n}$ represents the photon number operator.

The displacement receiver consists of beam splitters, 
displacements, on/off detectors w/o feedforward. 
The beam splitter 
operation $\hat{B}(R)$ combines and splits 
two input coherent states $|\beta\rangle$ 
and $|\gamma\rangle$ as 
\beqa
\hat B (R) \ket{\beta} \ket{\gamma}
& = & \ket{ \sqrt{1-R} \beta + \sqrt{R} \gamma }
\nonumber\\ &&
\otimes
\ket{ -\sqrt{   R} \beta + \sqrt{1-R} \gamma } , 
\eeqa
where its geometric configuration is 
illustrated in the inset of Fig.~\ref{Scheme_3PSK_2port}. 
Displacement operation $\hat{D}(\gamma)$ shifts 
the amplitude of coherent state as 
$\hat{D}(\gamma)|\beta\rangle = |\beta+\gamma\rangle$. 
It is well known that the displacement operation
is implemented by combining the signal and a local oscillator
via a highly transmissive beam splitter
(for example, see \cite{Tsujino2011_Q_Receiver_BPSK}).

On/off detector is a photon detection device observing only 
zero or non-zero photons. 
The on/off detector is described by a set of operators, 
\beqa\label{Imperfect_on_off_detector}
\hat{\Pi}_\mathrm{off}
&=&e^{-\nu}\displaystyle\sum^{\infty}_{n=0}(1-\eta)^n\ket{n}\bra{n}\;,
\label{eq:3}\\
\hat{\Pi}_\mathrm{on}&=&\hat{I}-\hat{\Pi}_\mathrm{off}\;,
\label{eq:4}
\eeqa
where $\nu$ is the dark count probability
and $\eta$ is the detection efficiency.
The probability of finding an off-signal when detecting $\ket{\alpha_m}$
is given by
\begin{equation}\label{P_off}
P_\mathrm{off}=
\bra{\alpha_m} \hat{\Pi}_\mathrm{off} \ket{\alpha_m}
=e^{-\nu-\eta\alpha^2}\;.
\end{equation}
%
\subsection{Ternary PSK signals: $M=3$}\label{Sect:2a}
\begin{figure}[t]
\begin{center}
\includegraphics[width=0.9\linewidth]
{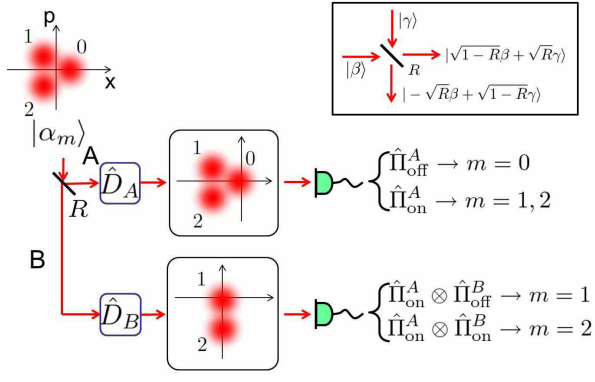}
\caption{Displacement receiver with two-port detection structure
without feedforward operations for the 3PSK case.
Inset represents the definition of the beam splitting operation for 
coherent states. See the text for details. 
}
\label{Scheme_3PSK_2port}
\end{center}
\end{figure}

%
The structure of the receiver for ${M=3}$ is depicted in
Fig.~\ref{Scheme_3PSK_2port}$\,$.
The basic operation principle follows the same idea
lying behind the Kennedy's receiver \cite{Kennedy73},
where
BPSK signals are firstly displaced,
such that one of the two signals becomes the vacuum state
(signal nulling),
and then
they are discriminated by means of an on/off detector.
For an ideal photon detector,
the vacuum state is always determined with no error,
while mis-detection may occur on the other state.
For multiple PSK signals,
we can extend the same basic principle
by nulling constellation symbols.

In  Fig.~\ref{Scheme_3PSK_2port},
the optical signal is split into two branches A and B
via a beam splitter having reflectance $R$. 
After the beam splitting, 
On branch A,
the signal is displaced by $\hat D_A(-\sqrt{R}\,\alpha_0)$
while on branch B by $\hat D_B(-\sqrt{1-R}\,\alpha_1)$.
As a result,
we can see the system as being made up of the composition of
two separable states,
leading to the following possible overall two-mode states
\begin{equation}
\begin{array}{llll}
\ket{\Psi_0}_{AB}
=\ket{0}_A \otimes \ket{{\sqrt {1-R}}(\alpha_0-\alpha_1)}_B\quad,
\\
\ket{\Psi_1}_{AB}
=\ket{\sqrt{R}(\alpha_1-\alpha_0)}_A \otimes \ket{0}_B
\quad,\\
\ket{\Psi_2}_{AB}
=\ket{\sqrt{R}(\alpha_2-\alpha_0)}_A
\otimes \ket{{\sqrt {1-R}}(\alpha_2-\alpha_1)}_B\quad,
\end{array}
\end{equation}
where, to simplify the description, we assume that the phase shift due to the beam splitter is compensated by a phase shifter.
 
By regarding the vacuum and non-vacuum states
as the ``on'' and ``off'' signals, respectively,
and by viewing the signals on branch A and B as couples,
the above states can be referred to as
$\ket{\Psi_0}_{AB}\rightarrow$(off, on),
$\ket{\Psi_1}_{AB}\rightarrow$(on, off),
and
$\ket{\Psi_2}_{AB}\rightarrow$(on, on).
Then, recalling \eqref{eq:3} and \eqref{eq:4},
a straightforward decision rule can be given
through the definition of the following operators
\begin{equation}
\begin{array}{llll}
\hat \Pi_0
=\hat \Pi_\mathrm{off}^A \otimes \hat \Pi_\mathrm{on}^B\;,
\\
\hat \Pi_1
=\hat \Pi_\mathrm{on}^A \otimes \hat \Pi_\mathrm{off}^B\;,
\\
\hat \Pi_2
=\hat \Pi_\mathrm{on}^A \otimes \hat \Pi_\mathrm{on}^B\;,
\\
\hat \Pi_3
=\hat \Pi_\mathrm{off}^A \otimes \hat \Pi_\mathrm{off}^B\;,
\end{array}
\end{equation}
where $\hat\Pi_3$ represents the (off, off) case.

The channel matrix $P(j\vert i)=\bra{\Psi_i}\hat\Pi_j\ket{\Psi_i}$
is then composed by the following elements
\begin{equation}
\begin{array}{llll}
P(0|0)
=e^{-\nu} (1-e^{-\nu-3\eta (1-R) \alpha^2})
\\
P(1|0)
=(1-e^{-\nu})e^{-\nu-3\eta (1-R) \alpha^2}
\\
P(2|0)
=(1-e^{-\nu})(1-e^{-\nu-3\eta (1-R) \alpha^2})
\\
P(3|0)
=e^{-\nu}e^{-\nu-3\eta (1-R) \alpha^2}
\\

P(0|1)
=e^{-\nu-3\eta R \alpha^2}(1-e^{-\nu})
\\
P(1|1)
=(1-e^{-\nu-3\eta R \alpha^2})e^{-\nu}
\\
P(2|1)
=(1-e^{-\nu-3\eta R \alpha^2})(1-e^{-\nu})
\\
P(3|1)
=e^{-\nu-3\eta R \alpha^2}e^{-\nu}
\\

P(0|2)
=e^{-\nu-3\eta R \alpha^2}(1-e^{-\nu-3\eta (1-R) \alpha^2})
\\
P(1|2)
=(1-e^{-\nu-3\eta R \alpha^2})e^{-\nu-3\eta (1-R) \alpha^2}
\\
P(2|2)
=(1-e^{-\nu-3\eta R \alpha^2})(1-e^{-\nu-3\eta (1-R) \alpha^2})
\\
P(3|2)
=e^{-\nu-3\eta R \alpha^2}e^{-\nu-3\eta (1-R) \alpha^2}
\\
\end{array}\label{eq:8}
\end{equation}

Following a maximum likelihood criterion,
we can associate to any outcome
a symbol estimate $\hat{\alpha}$ as follows:

\begin{equation}
\begin{array}{llll}
\mathrm{(off, on)}&\rightarrow&\hat{\alpha}=\alpha_0\;,
\\
\mathrm{(on, off)}&\rightarrow&\hat{\alpha}=\alpha_1\;,
\\
\mathrm{(on, on)}&\rightarrow&\hat{\alpha}=\alpha_2\;,
\\
\mathrm{(off, off)}&\rightarrow&\hat{\alpha}=\alpha_0\;, &
   \,\mathrm{if}\;R\geq 1/2\;,
\\
\mathrm{(off, off)}&\rightarrow&\hat{\alpha}=\alpha_1\;, &
   \,\mathrm{if}\;R< 1/2\;.
\end{array}
\end{equation}


\begin{figure}[h]
\centering
\subfigure
{
\includegraphics[width=0.95\linewidth]
{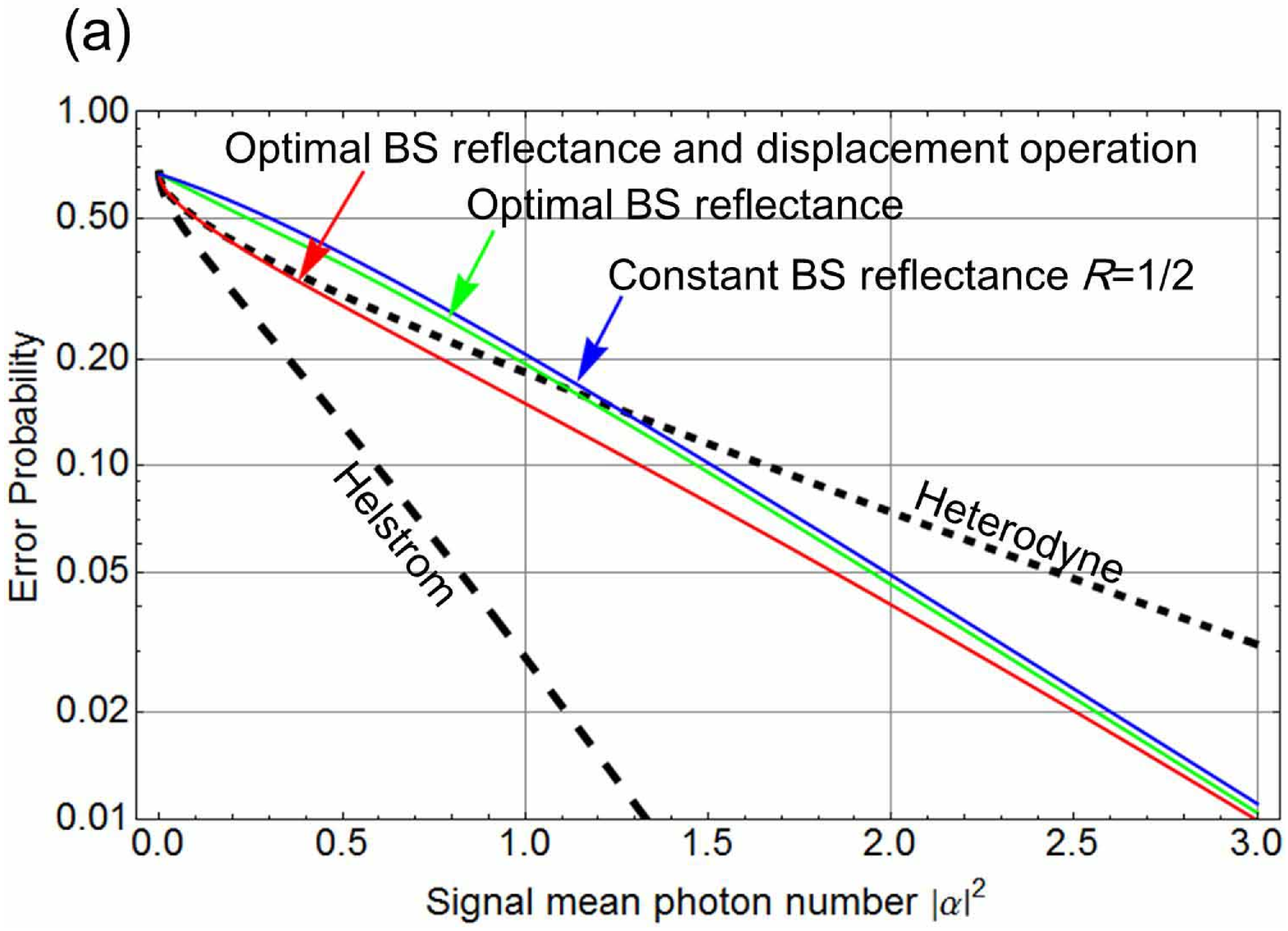}
\label{Fig3a}
}
\subfigure
{
\includegraphics[width=0.95\linewidth]
{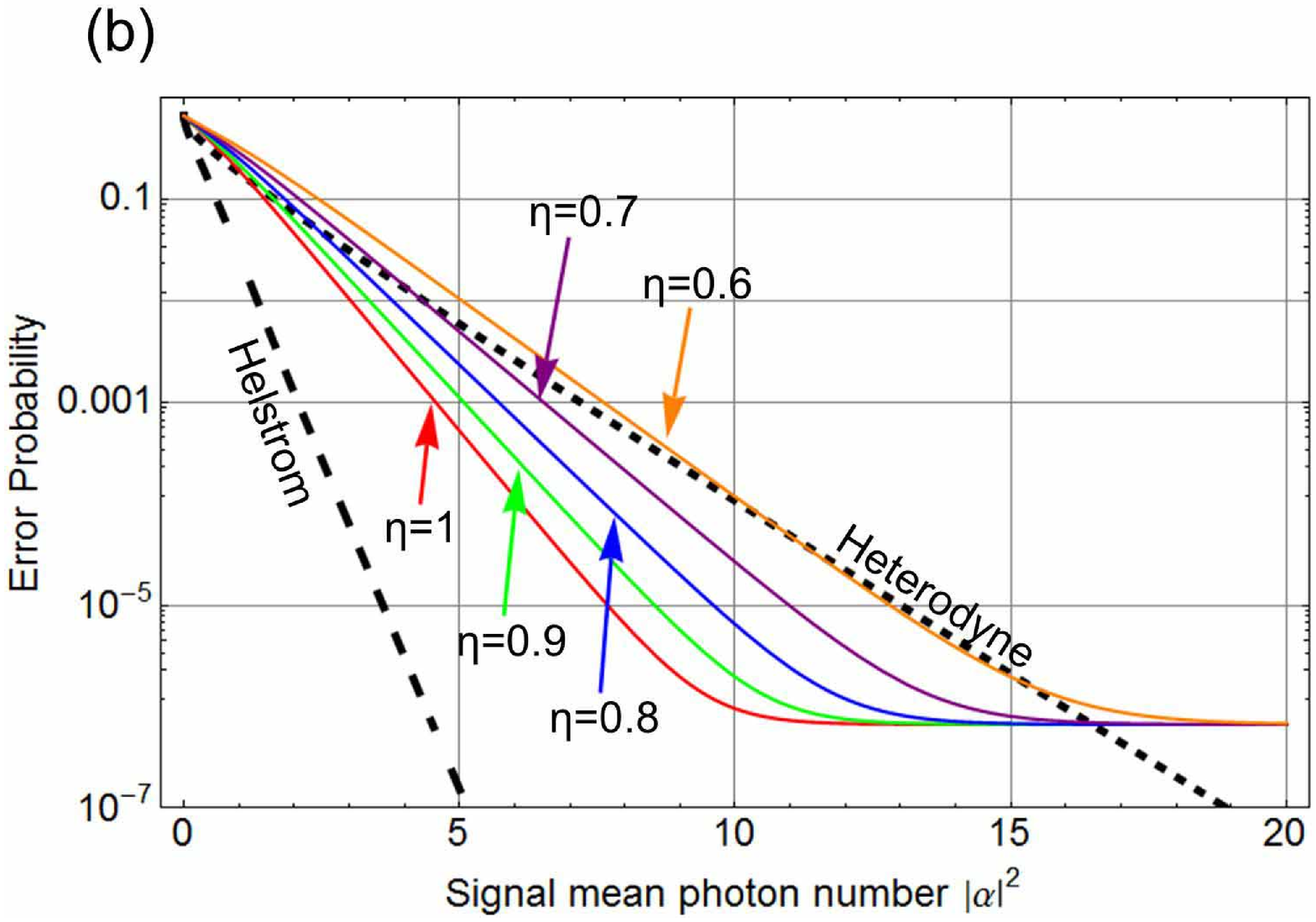}
\label{Fig3b}
}
\caption{
Average error rate for 3PSK signal discrimination without
applying feedforward.
(a) Equal beam splitting and exact nulling (blue line),
optimized beam splitting and exact nulling (green line), and
optimized beam splitting and displacements (red line).
$\eta=1$ and $\nu=0$.
(b) Error performance for different values of $\eta$. $\nu=10^{-6}$.
In both figures,
the black dashed and dotted lines represent the Helstrom and the
heterodyne limits, respectively.
}
\label{Fig3}
\end{figure}


Hence, for $R\geq 1/2$, the average error rate is given by
\begin{equation}
\begin{split}
P_e=
& 1- \frac{1}{3} \sum_{m=0}^2 P(\hat{\alpha}=\alpha_m | \alpha_m)
\\
=
& 1 - \frac{1}{3} \left\{P(0|0) + P(3|0) +P(1|1)+P(2|2)\right\}.
\end{split}
\label{eq:9}
\end{equation}
We note that when ``off'' is the outcome on branch A,
we decide for 0, regardless of the outcome on branch B.
The analysis can also be easily turned to the case ${R<1/2}$.

The error rates derived from Eq.~(\ref{eq:9}) are plotted in
Fig.~\ref{Fig3}(a).
The blue and green lines are obtained, respectively, for fixed $R=1/2$ and
for numerically optimized $R$ for any given value of $\alpha$.
The performance difference between the two setups is small and, for signals with $|\alpha|^2 > 2$, the proposed receiver remarkably outperforms the heterodyne limit.
In the weak coherent state region,
the receiver performance can be further improved
by optimizing the amount of the displacements
$\hat D_A(\cdot)$ and $\hat D_B(\cdot)$ (i.e., not the exact nulling)
as indicated by the red line. Displacement optimization was discussed in
\cite{TakeokaSasaki2008_DisplacementRec_GaussianLimit,
Wittmann2008_PRL_BPSK,Tsujino2010_OX_OnOff,Tsujino2011_Q_Receiver_BPSK, ADP11} for BPSK signals and in \cite{Guha2011_JMO} for the pulse position modulation. We observe that an additional gain can be obtained
in the weak signal region. In Fig.~\ref{Fig3}(b) we plot the error rate with optimized $R$
and exact nulling assuming imperfect on/off detectors having $\nu=10^{-6}$ and different values for $\eta$. We note that it is possible to outperform the heterodyne limit even with moderate detection efficiency.
For example, the TES developed
in \cite{Lita2008_OX_TES_DE95,Fukuda2009_Metrologia_TES}
already reached $\eta=90\%$ and $\nu=10^{-6}$ and thus
the sub-SQL receiver could be realized with currently available technology.

%
\subsection{Quaternary PSK signals: $M=4$}\label{Sect:2b}
%

\begin{figure}[t]
\begin{center}
\includegraphics[width=0.9\linewidth]
{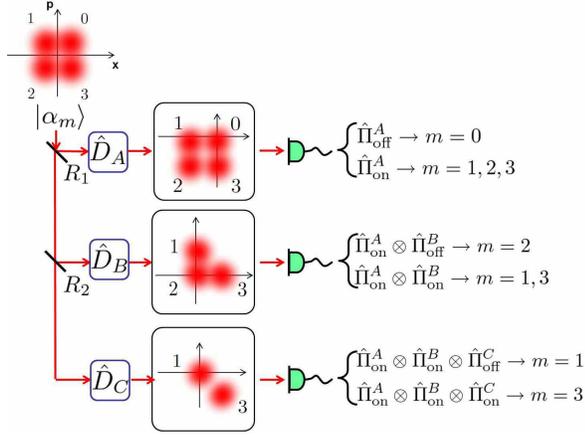}
\caption{Displacement receiver with three-port detection structure
without feedforward operations for the 4PSK case.}
\label{Fig4}
\end{center}
\end{figure}

For the 4PSK signal
we consider the three ports scheme depicted in
Fig.~\ref{Fig4}.
The input signal is split into three branches A, B, and C
by means of two beam splitters with reflectance $R_1$ and $R_2$,
respectively.
Based on the three outcomes,
the optimal decision rule can be pursued
by following a similar approach as for the 3PSK modulation.
It turns out that, different strategies can be adopted
depending on the value of $\alpha$ and selection of $R_1$ and $R_2$.
By an analytical and numerical study
we found that the following straightforward strategy can be employed
without any noteworthy performance degradation.

On the first branch A,
the signal $\ket {\sqrt{R_1}\,{\alpha _m}}_A$
is displaced by $\hat{D}_A (-\sqrt{R_1}\,\alpha_0)$
and it is detected by an on/off detector.
If the result is ``off'', then the symbol estimate
is taken as $\hat{\alpha}=\alpha_0$,
otherwise the results on the successive branches are considered.
At this stage,
the probability of correct decision for symbol $\alpha_0$ results
\begin{eqnarray}
P(\hat{\alpha}=\alpha_0|\alpha_0)=e^{-\nu }\;.
\label{eq:10}
\end{eqnarray}

If the result is ``on'' on branch A,
we discharge the hypothesis of symbol $\alpha_0$.
On branch B,
the signal $\ket{\sqrt{(1-R_1)R_2}\,\alpha _m}_B$ is displaced by
$\hat{D}_B(-\sqrt{(1-R_1)R_2}\,\alpha_2)$.
If the result is ``off'', then the estimate
is taken as $\hat{\alpha}=\alpha_2$,
otherwise the result on the next branch is considered.
The probability of correct decision for $\alpha_2$,
is given by the product of the probabilities of the events:
having ``on'' on branch A and having ``off'' on branch B, that is
\begin{eqnarray}
P(\hat{\alpha}=\alpha_2|\alpha_2)=\left(1-e^{-\nu -4\eta R_1 \alpha ^{2}}\right)e^{-\nu}\;.
\end{eqnarray}

Next, if the result is ``on'' on branch B,
then we attempt to distinguish
between $\alpha_1$ and $\alpha_3$ on the last branch C.
So the signal $\ket{\sqrt{(1-R_1)(1-R_2)}\,\alpha _m}_C$
is displaced by $\hat{D}_C(-\sqrt{(1-R_1)(1-R_2)}\,\alpha _1)$,
and if the outcome is ``off''
we decide $\hat{\alpha}=\alpha_1$,
otherwise $\hat{\alpha}=\alpha_3$.
The probabilities of correct decision results
\begin{eqnarray}
P(\hat{\alpha}=\alpha_1|\alpha_1)
&=&\left(1-e^{-\nu -2\eta R_1 \alpha ^{2}}\right)
\nonumber
\\
&\times&\left(1-e^{-\nu -2\eta (1-R_1) R_2 \alpha ^{2}}\right)
e^{-\nu}\;,
\\
P(\hat{\alpha}=\alpha_3|\alpha_3)
&=&\left(1-e^{-\nu -2\eta R_1 \alpha ^{2}}\right)
\nonumber
\\
&\times&\left(1-e^{-\nu -2\eta (1-R_1) R_2 \alpha ^{2}}\right)
\nonumber
\\
&\times&\left(1-e^{-\nu -4\eta (1-R_1) (1-R_2)\alpha ^{2}}\right)\;.
\end{eqnarray}

Therefore, the average error rate is given by
\begin{eqnarray}
P_e=1-\frac{1}{4}\sum_{m=0}^{3} P(\hat{\alpha}=\alpha_m|\alpha_m)\;.
\end{eqnarray}

\begin{figure}[t]
\centering
{
\includegraphics[width=0.95\linewidth]
{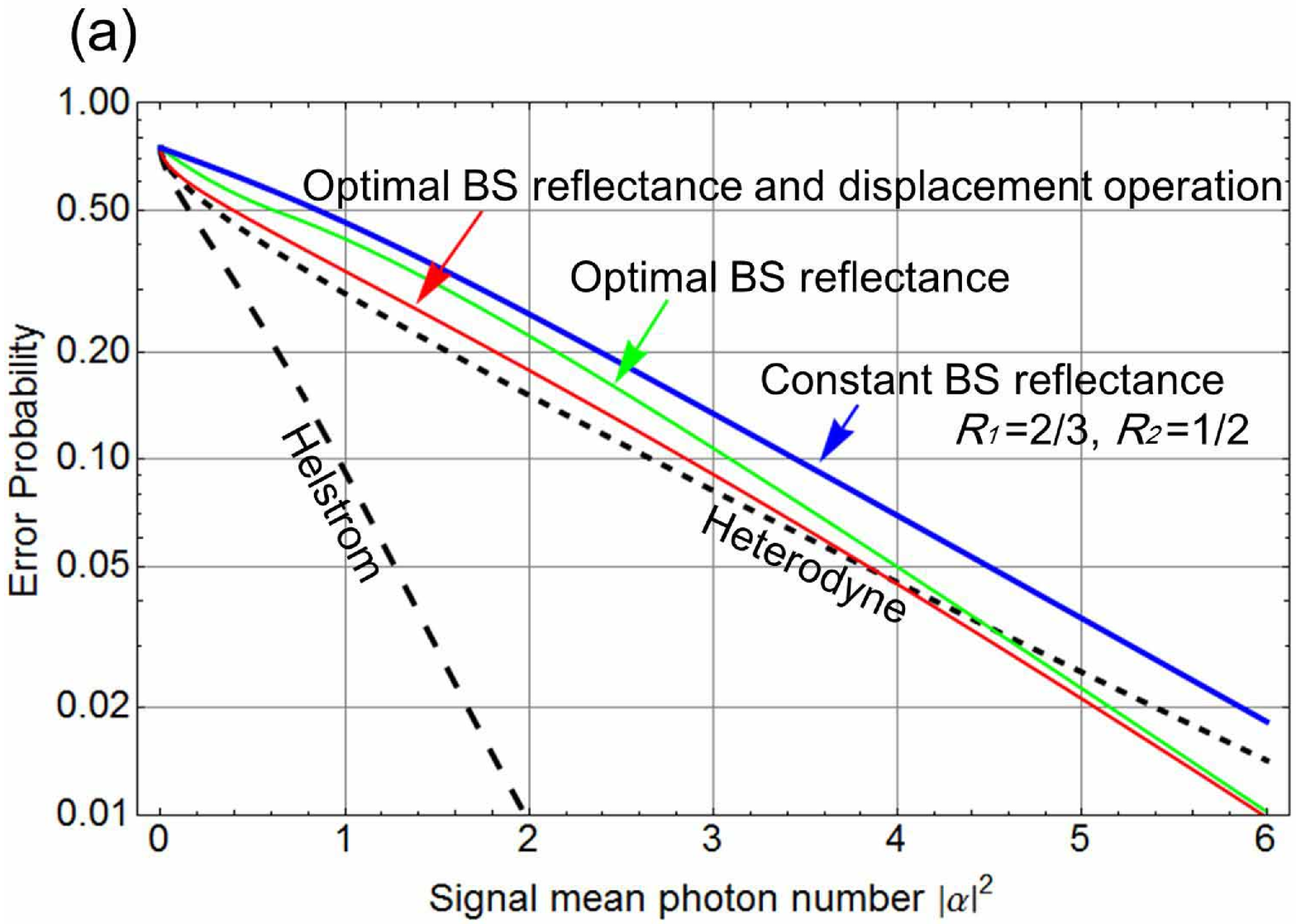}
\label{Fig5a}
}
{
\includegraphics[width=0.95\linewidth]
{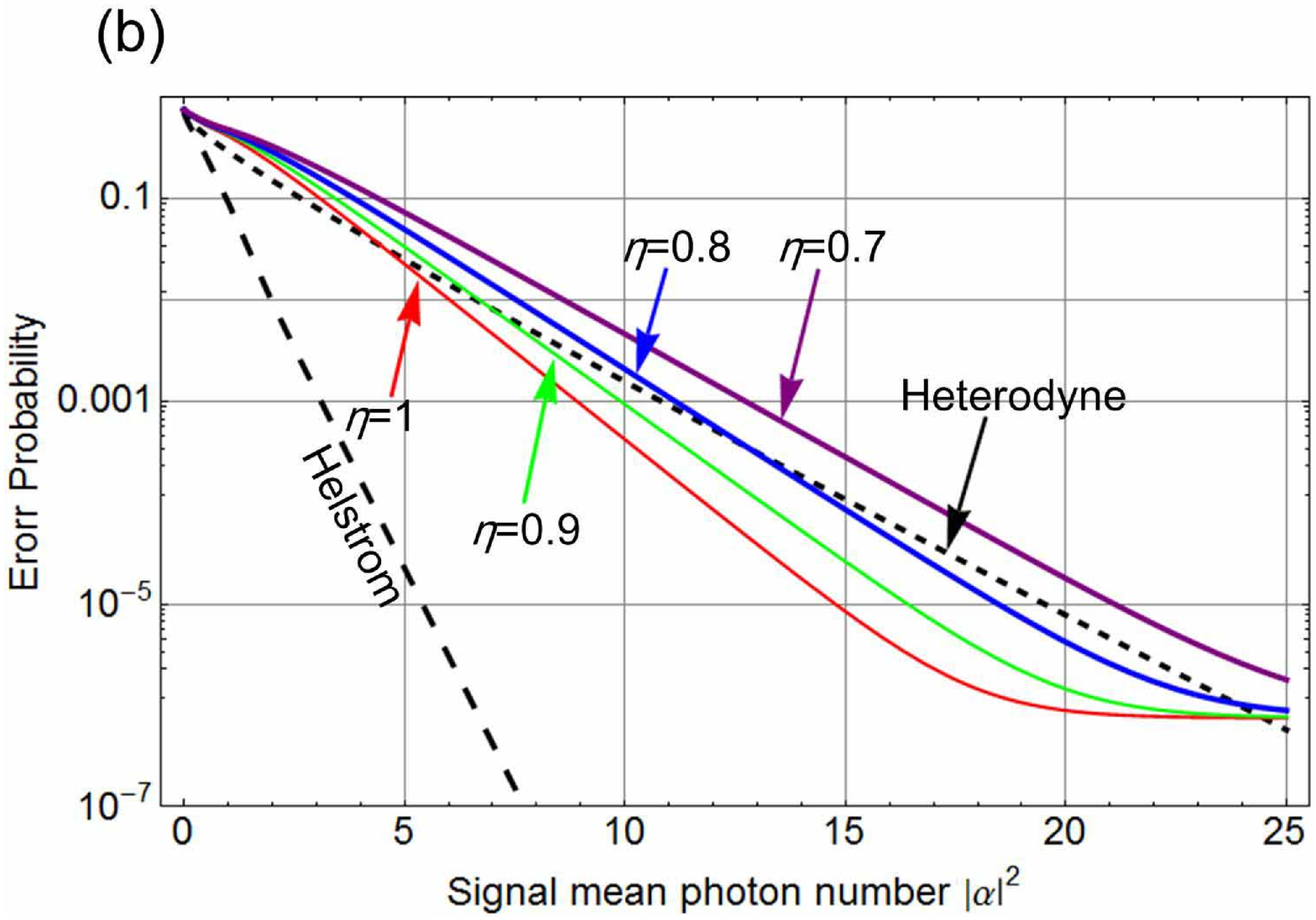}
\label{Fig5b}
}
\caption{
Average error rate for 4PSK signal discrimination without
applying feedforward.
(a) Equal beam splitting and exact nulling (blue line),
optimized beam splitting and exact nulling (green line), and
optimized beam splitting and displacements (red line).
$\eta=1$ and $\nu=0$.
(b) Error performance for different values of $\eta$. $\nu=10^{-6}$.
In both figures,
the black dashed and dotted lines represent the Helstrom and the
heterodyne limits, respectively.}
\label{Fig5}
\end{figure}

Figure \ref{Fig5}(a) reports the resulting error rates
with equal beam splitting ($R_1=2/3$, $R_2=1/2$, the blue line),
and optimized
$R_1$ and $R_2$ w/o the optimization of the displacements
(the green and red lines, respectively)
in comparison with the heterodyne limit and the Helstrom bound.
In contrast to the receiver for 3PSK signals, for the 4PSK case the optimization of the reflectances is crucial to provide better performance than the heterodyne limit, while the optimization of the displacements
is less effective. The effect of detector imperfections are illustrated in
Fig.~\ref{Fig5}(b). We note that the requirement on detector efficiency is quite severe and the expected gain with respect to the heterodyne limit is not as remarkable as for the 3PSK case.

%
\section{Displacement receiver with feedforward}\label{Sect:3}
%

\begin{figure}[t]
\begin{center}
\includegraphics[width=0.8\linewidth]
{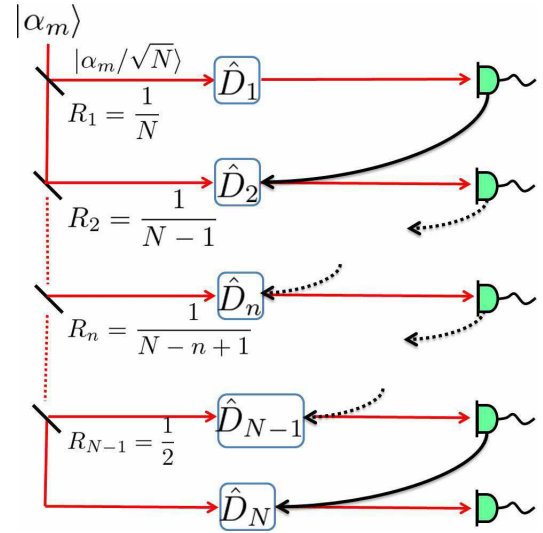}
\caption{
The displacement receiver consisting of $N$-step
feedforward operations.
}
\label{Fig6}
\end{center}
\end{figure}

In this section, we discuss the displacement receiver
employing feedforward operations.
The schemes discussed in Sect.~\ref{Sect:2}
were composed of a fixed number $M-1$ of branches
dependent on the number of signals $M$.
Hereinafter,
we consider a generalization
where the incoming signal is split into a generic number
$N\geq M-1$ of branches as shown in Fig.~\ref{Fig6}.


The reflectance of the displacement at the $n$th branch,
$n=1,2,\ldots N-1$, is fixed to $R_n=1/(N-n+1)$,
so that the signal intensity is the same on each branch.
In other words,
we obtain $N$ copies of weaker state $\k(\alpha_m/\sqrt{N})$
of the incoming signal $\k(\alpha_m)$
(this could also be done in the time domain if convenient).
We also assume that the value of the displacement at the $n$th branch,
$\hat{D}_n(\cdot)$,
can be set once the outcome on the $(n-1)$th branch is obtained.
Final decision is performed considering the outcomes
obtained on the $N$ different branches.

In the following, we detail the detection strategy for 3PSK and 4PSK signals.

%
\subsection{Ternary PSK signals: $M=3$}\label{Sect:2a}
%

The use of feedforward operations open to refine the decision process. In fact, at the first step $n=1$ we apply the same displacement $\hat D_1(-\alpha_0/\sqrt{N})$ as for the schemes considered in the previous section, however, if an ``off" signal is detected we do not definitely conclude that $\hat{\alpha}=\alpha_0$, but we just keep applying the same displacement also on the successive step to further validate our decision.

\begin{figure}[th!]
\begin{center}
\includegraphics[width=1.0\linewidth]
{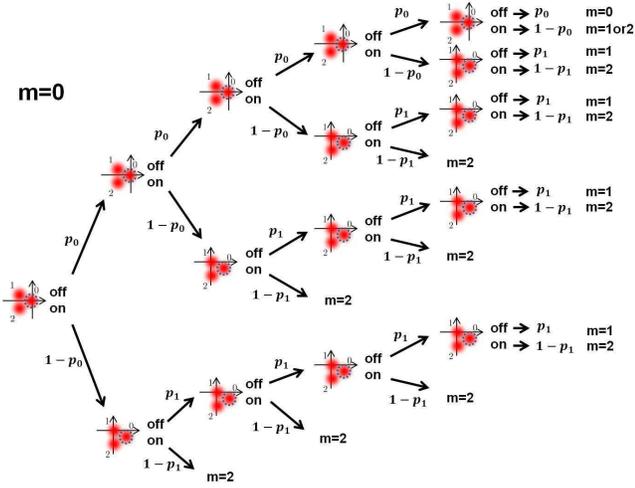}
\caption{Feedforward tree for the 3PSK case
where the input is $\alpha_0$ ($m=0$) and $N=5$.}
\label{3PSK_tree_m0}
\end{center}
\end{figure}

Fig.~\ref{3PSK_tree_m0} depicts an example of the feedforward tree for $N=5$ with signal input $\alpha_0$.
The probability of having an ``off'' outcome at the first step is
\begin{eqnarray}
p_0=\mathrm{e}^{-\nu}\;. \label{eq17}
\end{eqnarray}
Hence, because of the dark counts, with probability $1-p_0$, the result ``on" may occur and, erroneously, the receiver try to discriminate between  symbols $m=1$ and $m=2$. Therefore, in the second step,  the displacement $\hat D_2(-\alpha_1/\sqrt{N})$ is applied with the aim of testing hypothesis $m=1$. Then, if an ``off'' is detected,
we maintain the same displacement and further proceed to the third step;
such an event occurs with probability
\begin{eqnarray}
p_1=\mathrm{e}^{-\nu-\frac{3\eta\alpha^2}{N}}\;. \label{eq18}
\end{eqnarray}
Otherwise, if the detection returns an ``on'' signal,
then we just erroneously decide for the signal $\hat{\alpha}=\alpha_2$.
Similar operations are repeated up to the $N$th step.

The resulting decision rule is conveniently summarized as follows.
When all the detectors output ``off",
we decide for $\hat{\alpha}=\alpha_0$. If only one ``on'' is detected in the first $N-1$ steps and an ``off'' is detected in the last step, then $\hat{\alpha}=\alpha_1$. If only one ``on'' signal is detected at the last step, the estimate is randomly made between $\hat{\alpha}=\alpha_1$ and $\hat{\alpha}=\alpha_2$. Finally, if at least two ``on'' signals are detected, then $\hat{\alpha}=\alpha_2$.

The probabilities of correct decision are then given by
\begin{eqnarray}
P(\hat{\alpha}=\alpha_0\vert \alpha_0)&=&p_{0}^N\;,
\label{eq19}
\\
P(\hat{\alpha}=\alpha_1\vert \alpha_1)&=&\sum_{t=0}^{N-2}p_{1}^t \left( 1-p_{1} \right) p_{0}^{N-1-t}
\nonumber
\\
&&+p_{1}^{N-1}\left( 1-p_{1} \right) \times \frac{1}{2}\;,
\label{eq20}
\\
P(\hat{\alpha}=\alpha_2 \vert \alpha_2)&=&\sum_{t=1}^{N-2}p_{1}^t  \left( 1-p_{1} \right)
\sum_{s=0}^{N-2-t} p_{1}^s  \left(1-p_{1} \right)
\nonumber
\\
&&
+ \sum_{t=0}^{N-2}\left(1-p_{1} \right) p_{1}^t \left(1-p_{1} \right)
\nonumber
\\
&&+p_{1}^{N-1} \left( 1-p_{1} \right) \times \frac{1}{2}\;.
\nonumber
\\
\label{eq21}
\end{eqnarray}

The average error rate is then equal to
\beq
P_e=1-\frac{1}{3}\sum_{m=0}^{2} P(\hat{\alpha}=\alpha_m \vert \alpha_m)\;.
\eeq

Assuming zero dark counts ($\nu=0$)
the above equations simplify as
\begin{eqnarray}
P(\alpha_0 \vert \alpha_0)&\!\!=\!\!&1\;,
\\
P(\alpha_1 \vert \alpha_1)&\!\!=\!\!&1\!-\!\frac{1}{2}\left( p_{1}^N +p_{1}^{N-1} \right)\;,
\\
P(\alpha_2 \vert \alpha_2)&\!\!=\!\!&1\!-\!\frac{1}{2} \left[ \left( 2 N\!-\!1\right) p_{1}^{ N-1}
 \!-\!\left( 2N\!-\!3 \right) p_{1}^N \right],
\end{eqnarray}
and
\begin{equation}\label{eq:Pe3pskidric}
P_e
=
\frac{1}{3}\mathrm{e}^{-3 \eta \alpha^2} \left[ 2+ N \left( \mathrm{e}^{+3 \eta \frac{\alpha^2}{N}}-1 \right) \right].
\end{equation}
In the limit of $N\rightarrow\infty$, we obtain
\begin{equation}\label{P_e:3PSK:N=infty}
P_e^{\infty}
=
\frac{1}{3} \mathrm{e}^{-3 \eta \alpha^2 } \left( 2+3\eta\alpha^2 \right).
\end{equation}

The performance assuming ideal on/off detectors are shown in
Fig.~\ref{3PSK_BER_N_infinity2}.
The error rate noticeably decreases with the increasing of $N$. Most of the gain is achieved with just $N=5$, and with $N=10$ the performance gets very close to the asymptotical bound \eqref{P_e:3PSK:N=infty}. For a sufficiently high signal intensity (such as $\alpha^2\gg 2/(3\eta)$) the bound \eqref{P_e:3PSK:N=infty}, for $\eta=1$, approximates as
\beq\label{P_e:3PSK:N=infty:scaling}
P_e^{\infty}
\sim
\alpha^2 \mathrm{e}^{-3 \alpha^2 }\,.
\eeq

For 3-PSK signal the Helstrom bound is given by
\cite{Ban97}
\beq
P_{e,Hel}
=
1-\frac{1}{9}\left( \sum_{m=0}^2 \sqrt{\lambda_m} \right)^2\;,
\eeq
where
\begin{equation}\label{Lambda_3PSK}
\begin{array}{llll}
\lambda_0&=&1+2\kappa_c\;,
\\
\lambda_1&=&1-\kappa_c+\sqrt{3}\kappa_s\;,
\\
\lambda_2&=&1-\kappa_c-\sqrt{3}\kappa_s\;,
\end{array}
\end{equation}
with
\begin{equation}
\begin{array}{llll}
\kappa_c&=&\exp\left(-{3\over2}\alpha^2\right)
           \cos\left({\sqrt{3}\over2}\alpha^2\right)\;,
\\
\kappa_s&=&\exp\left(-{3\over2}\alpha^2\right)
           \sin\left({\sqrt{3}\over2}\alpha^2\right)\;,
\end{array}
\end{equation}
and for large values of $\alpha^2$ we find
\beq\label{P_e_Helstrom:3PSK:N=infty:scaling}
P_{e,Hel}
\sim
\frac{1}{2} e^{-3 \alpha^2 }.
\eeq
Therefore, from the comparison between \eqref{P_e:3PSK:N=infty:scaling} and \eqref{P_e_Helstrom:3PSK:N=infty:scaling}, we note that the asymptotical performance gap between the feedforward receiver and the Helstrom bound depends on the signal intensity $\alpha^2$.

Fig.~\ref{3PSK_BER_N_step_DE90DK-6} points out the
impact of imperfect detectors on the system error rate for different values of $N$ ($\eta=90$\%, $\nu=10^{-6}$).
We observe that for large $N$ the effect of
the dark counts accumulate and seriously degrades
the performance.
For example, for $\alpha^2>10$,
the simpler 2-port scheme proposed in Sect.~\ref{Sect:2}
attains better performance than the feedforward scheme.
The dependence on the detector efficiency is illustrated
in detail in Fig.~\ref{3PSK_BER_N10_DE_dependence} for $N=10$.
The figure shows that the gain due to the feedforward could
provide more tolerance to detector efficiency, in agreement with the results in \cite{Becerra_NIST_2011_MPSK_emulation_experiment}.
Finally, it should be noted that the optimization of
the displacements also works for the feedforward receivers
although the additional gain is small, see Appendix A.

{
\begin{figure}[t!]
\begin{center}
\includegraphics[width=0.95\linewidth]
{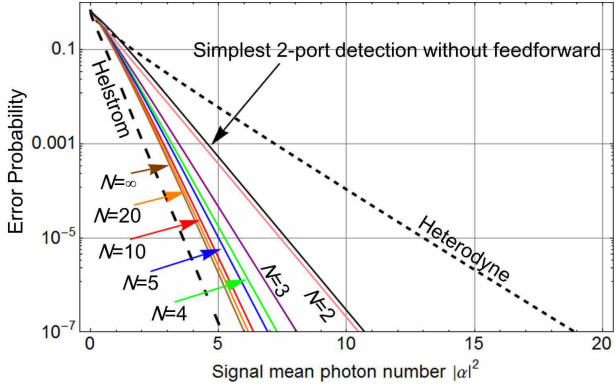}
\caption{
Average error rates for 3PSK signal discrimination with $N$-step feedforward operation with perfect detectors:
$\nu=0$ and $\eta=1$.
}
\label{3PSK_BER_N_infinity2}
\end{center}
\end{figure}

\begin{figure}[th!]
\begin{center}
\includegraphics[width=0.95\linewidth]
{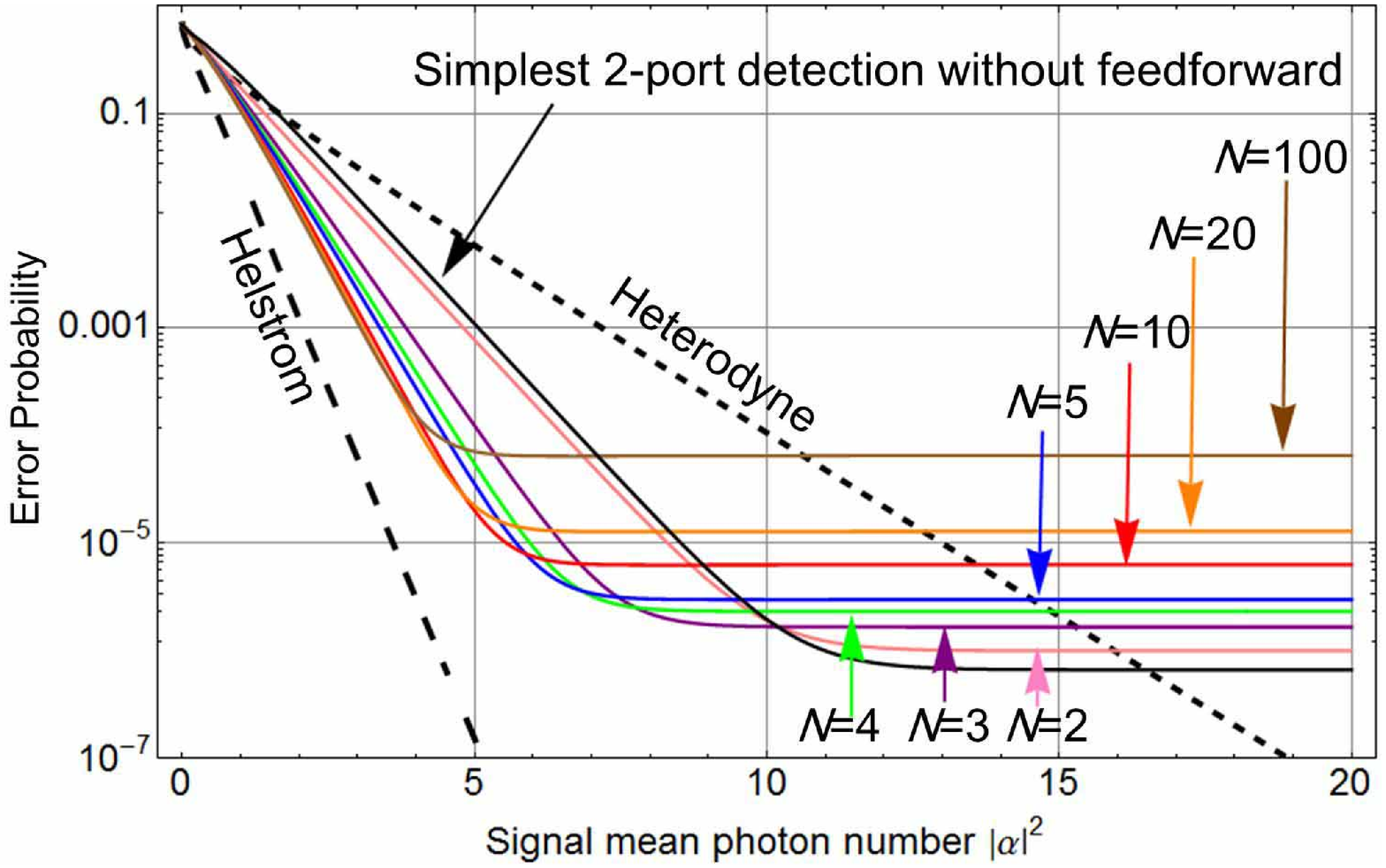}
\caption{
Average error rates for 3PSK signal discrimination with $N$-step feedforward operation and imperfect detectors:
$\nu=10^{-6}$ and $\eta=0.9$.
}
\label{3PSK_BER_N_step_DE90DK-6}
\end{center}
\end{figure}

\begin{figure}[th!]
\begin{center}
\includegraphics[width=0.95\linewidth]
{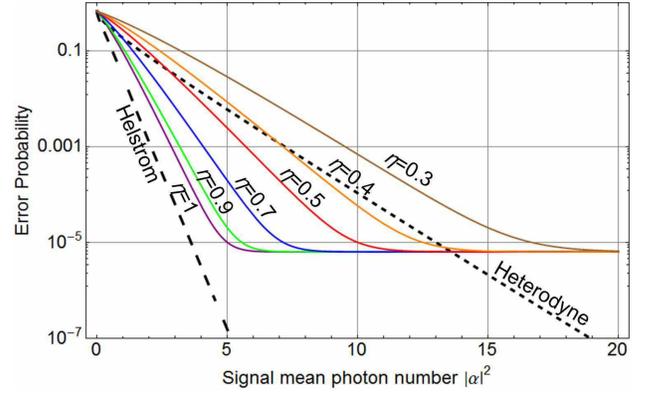}
\caption{
Average error rates for 3PSK signal discrimination with $N$-step feedforward operation for $N=10$, $\nu=10^{-6}$ and different values of $\eta$.
}
\label{3PSK_BER_N10_DE_dependence}
\end{center}
\end{figure}

%
\subsection{Quaternary PSK signals: $M=4$}\label{Sect:3b}
%

Similar arguments as in the previous section can be applied
to the 4PSK state discrimination.
In this case, at the first step, we have to define three different probabilities
of getting an ``off'' signal after displacement
$\hat D_1(-\alpha_0/\sqrt{N})$
\begin{eqnarray}
p_0&=&\mathrm{e}^{-\nu}\;, \label{eq22}
\\
p_1&=&\mathrm{e}^{-\nu-\frac{2\eta \alpha^2}{N} }\;, \label{eq23}
\\
p_2&=&\mathrm{e}^{-\nu-\frac{4\eta \alpha^2}{N} }\;. \label{eq24}
\end{eqnarray}
%
%
%
The probabilities of correct decision result
%
%
\begin{eqnarray}
P(0|0)
&=&p_{0}^{N}\;,
\label{eq25}
\\
P(1|1)
&=&\sum_{t=0}^{N-3}p_{1}^{t} \left(1-p_{1}\right)
\sum_{s=0}^{N-3-t} p_{1}^{s}  \left(1-p_1 \right) p_0^{N-2-t-s}
\nonumber
\\
&&
+\sum_{t=0}^{N-2}p_{1}^{t} \left( 1-p_{1} \right)
p_{1}^{N-2-t} \left(1-p_1\right) \times \frac{1}{2}
\nonumber
\\
&&
+p_1^{N-1} \left( 1-p_1\right)
\times \frac{1}{3}\;,
\label{eq26}
\\
P(2|2)
&=&\sum_{t=0}^{N-2} p_{2}^{t} \left( 1-p_{2} \right)
p_0^{N-1-t}
\nonumber
\\
&&+ p_{2}^{N-1} \left( 1-p_{2} \right)
\times \frac{1}{3}\;,
\label{eq27}
\\
P(3|3)
&=&\sum_{t=0}^{N-3} p_{1}^{t} \left( 1-p_{1} \right)
\sum_{s=0}^{N-3-t} p_1^{s} \left( 1- p_1 \right)
\nonumber
\\
&&\times
\sum_{u=0}^{N-3-t-s} p_2^{u} \left(1-p_2 \right)
\nonumber
\\
&&
+ \sum_{t=0}^{N-2} p_{1}^{t} \left( 1-p_{1} \right)
p_{1}^{N-2-t} \left( 1-p_1 \right)
\times \frac{1}{2}
\nonumber
\\
&&
+p_{1}^{N-1} \left( 1-p_{1} \right)
\times \frac{1}{3}\;.
\label{eq28}
\end{eqnarray}

To see the asymptotic behavior for $N$,
let us fix $\nu=0$ and then simplify the above equations as
\beqa
P(0|0)&=&1\;,
\\
P(1|1)&=&
1+\frac{1}{6}
\Bigl[ \left( 3 N-5\right) p_{1}^N -4 p_{1}^{N-1}
\nonumber
\\
&&-3\left( N-1\right) p_{1}^{N-2}
\Bigr]\;,
\\
P(2|2)
&=&1-\frac{1}{3}\left( p_{2}^{N} + 2 p_{2}^{N-1} \right)\;,
\\
P(3|3)
&=&1+\frac{1}{6}
\Bigl[ \left( 9 N-11\right) p_{1}^N
-\left( 6 N-8 \right) p_{1}^{N-1}
\nonumber
\\
&&-3 \left( N-1\right) p_{1}^{N-2}-6 p_{2}^{N-1}
\Bigr]\;.
\eeqa
%
The error rates are plotted in
Fig.~\ref{4PSK_BER_N_infinity2} for ideal on/off detectors.
The error rate is remarkably improved by increasing $N$,
especially up to $N \approx 10$.

In the limit of $N\rightarrow\infty$, we obtain
\begin{equation}\label{P_e:4PSK:N=infty}
P_e^{\infty}
=
\frac{1}{2} e^{-4 \eta \alpha^2 }
+
\frac{1+6\eta\alpha^2}{4} e^{-2 \eta \alpha^2 }\;.
\end{equation}
which is further simplified for large $\alpha^2$
and $\eta=1$ as
\beq\label{P_e:4PSK:N=infty:scaling}
P_e^{\infty}
\sim
\frac{3}{2}\alpha^2 e^{-2 \alpha^2}\;.
\eeq

For 4-PSK signals the Helstrom bound reads
\beq \label{eq:bound4psk}
P_{e,Hel}
=
1-\frac{1}{16}\left( \sum_{m=0}^3 \sqrt{\lambda_m} \right)^2\;,
\eeq
where
\begin{equation}\label{Lambda_4PSK}
\begin{array}{llll}
\lambda_0&=&2 e^{-\alpha^2}(\cosh\alpha^2 + \cos\alpha^2)\;,
\\
\lambda_1&=&2 e^{-\alpha^2}(\sinh\alpha^2 + \sin\alpha^2)\;,
\\
\lambda_2&=&2 e^{-\alpha^2}(\cosh\alpha^2 - \cos\alpha^2)\;,
\\
\lambda_3&=&2 e^{-\alpha^2}(\sinh\alpha^2 - \sin\alpha^2)\;.
\end{array}
\end{equation}
The bound \eqref{eq:bound4psk} scales for large $\alpha^2$ as
\beq\label{P_e_Helstrom:4PSK:N=infty:scaling}
P_{e,Hel}
\sim
\frac{1}{2} e^{-2 \alpha^2 }\;,
\eeq
which again implies that the difference between the feedforward receiver
and the Helstrom limit is related to the signal intensity through a multiplicative factor $\alpha^2$.

It is also worth noticing that the asymptotic performance of
our receiver given by Eq.~(\ref{P_e:4PSK:N=infty:scaling})
basically coincides with that of the Bondurant receiver \cite{Bondurant93}. 
The error rate of the Bondurant receiver converges to 
$P_{e,Bon} \sim \alpha^2 e^{-2\alpha^2}$ for large $\alpha^2$ 
which is the same as Eq.~(\ref{P_e:4PSK:N=infty:scaling}) 
except the lack of coefficient $3/2$. 
The difference is due to the fact that 
the ordering of the pulse nulling is not the same 
($0\to1\to2$ in \cite{Bondurant93} while we choose 
$0\to2\to1$).
Though our ordering is not optimal in the asymptotic limit, 
we numerically found that 
our ordering shows lower error rates than that in \cite{Bondurant93} 
for small $N$ and also for the receiver without feedforward.

\begin{figure}[t]
\begin{center}
\includegraphics[width=0.95\linewidth]
{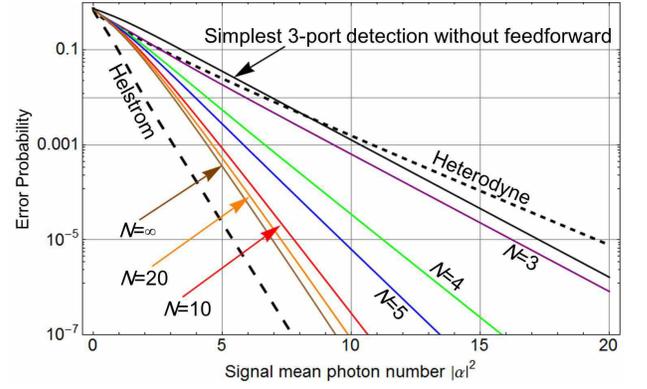}
\caption{
Average error rates for 4PSK signal discrimination with $N$-step feedforward operation with perfect detectors: $\nu=0$ and $\eta=1$.
}
\label{4PSK_BER_N_infinity2}
\end{center}
\end{figure}

The gap between the error rate of the feedforward receiver
and the Helstrom bound can be further reduced
by refining the feedforward rule
by adopting the maximization of the {\it a-posteriori} probabilities
as numerically demonstrated in
\cite{Becerra_NIST_2011_MPSK_emulation_experiment}.
We derive mathematical expression for this scheme in the 4PSK case
(see Appendix B) and report the error rate in
Fig.~\ref{4PSK_BER__BayesRule_N_step_DE100DK0}
for comparison with Fig.~\ref{4PSK_BER_N_infinity2}.

\begin{figure}[t]
\begin{center}
\includegraphics[width=0.95\linewidth]
{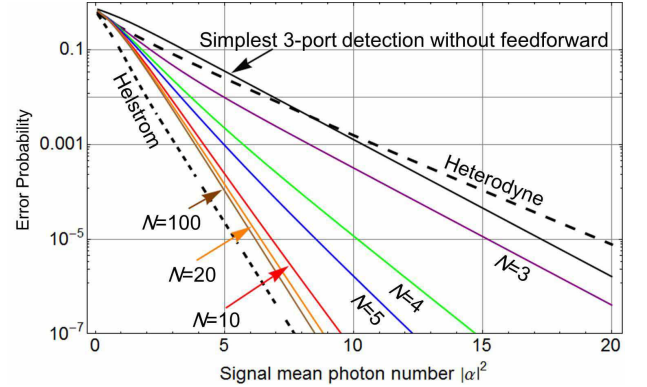}
\caption{Improved average error rates for 4PSK signal discrimination
obtained by refining the feedforward rule
with the maximization of {\it a--posteriori} probabilities.
Ideal detectors: $\nu=0$ and $\eta=1$.
}
\label{4PSK_BER__BayesRule_N_step_DE100DK0}
\end{center}
\end{figure}

Figure \ref{4PSK_BER_N_step_DE90DK-6} includes
the effect of the detector imperfections
($\eta=0.9$, $\nu=10^{-6}$)
into Fig.~\ref{4PSK_BER_N_infinity2}.
In contrast to the 3PSK case,
the scheme without any feedforward
cannot beat the heterodyne limit for $\alpha^2 < 20$.
It strongly suggests that
the feedforward would be essential to overcome the heterodyne limit
in practice.
For the dark count probability of $\nu=10^{-6}$,
$N=5\sim10$ would be a sensible choice.
Dependence on the detection efficiency is also highlighted
in Fig.~\ref{4PSK_BER_N10_DE_dependence} for $N=10$.

\begin{figure}[t]
\begin{center}
\includegraphics[width=0.95\linewidth]
{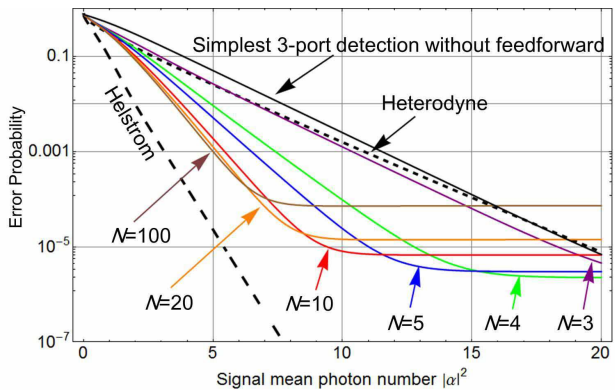}
\caption{
Average error rates for 4PSK signal discrimination with $N$-step feedforward operation and imperfect detectors:
$\nu=10^{-6}$ and $\eta=0.9$.
}
\label{4PSK_BER_N_step_DE90DK-6}
\end{center}
\end{figure}
%
%
\begin{figure}[H]
\begin{center}
\includegraphics[width=0.95\linewidth]
{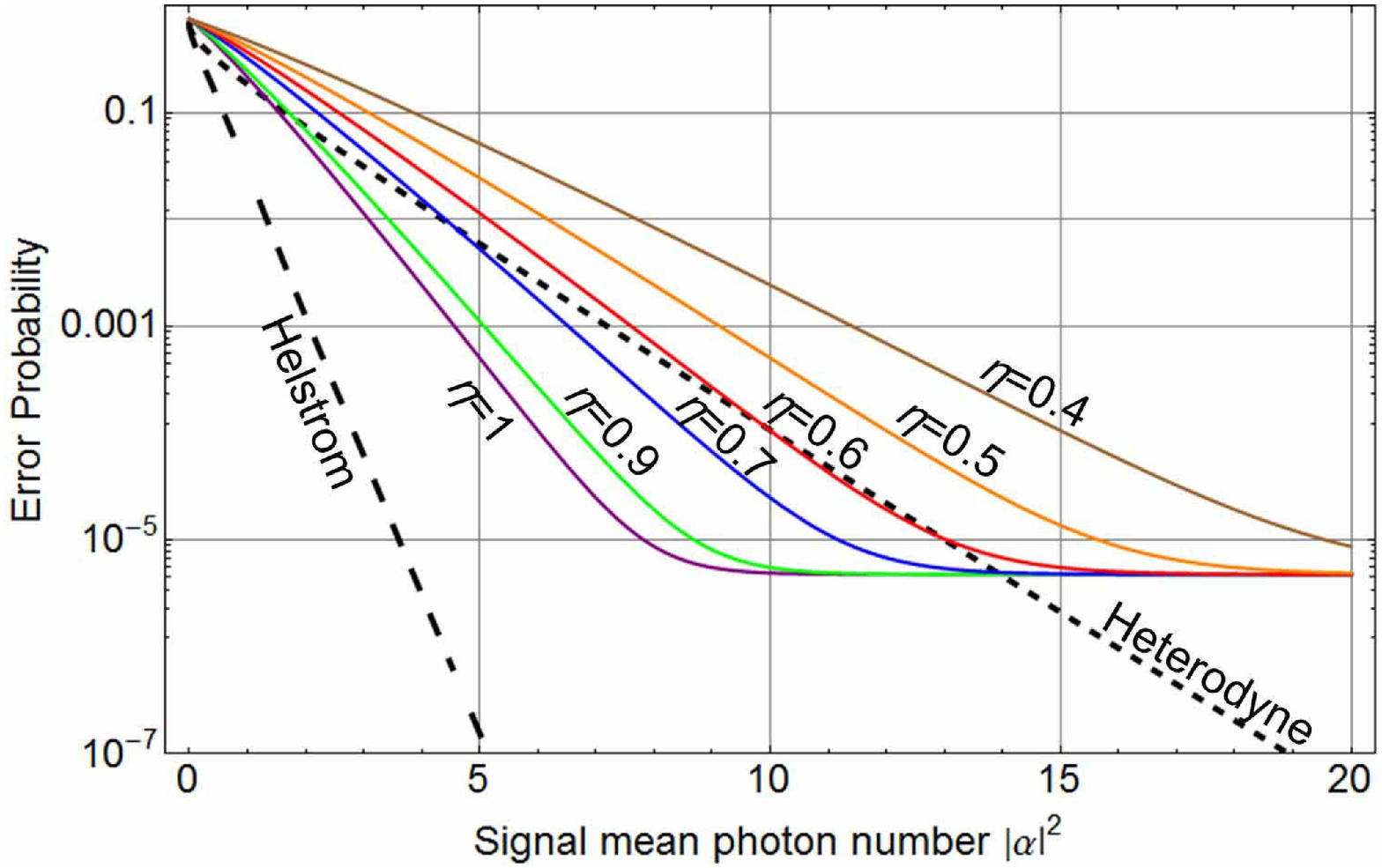}
\caption{
Dependence on the detection efficiency of the 4PSK signal
detection with feedforward. $N=10$ and $\nu=10^{-6}$.
}
\label{4PSK_BER_N10_DE_dependence}
\end{center}
\end{figure}
%
%
%
%
%
\section{Mutual Information}\label{Sect:4}
%

In this section we evaluate the mutual information
attained by the proposed displacement receivers.

The mutual information is related to the transmission efficiency
of reliable communication when coding techniques are employed.
Given the channel matrix of the transition probabilities $[P(y\vert x)]$
between input symbols $\{x\}$ and output symbols $\{y\}$, and the a-priori probabilities $\{P(x)\}$, the mutual information is given by
\cite{Shannon48,Gallager_book,CoverThomas_book}
\begin{equation}
I\left(X:Y\right)
=
\sum_x P\left(x\right)\sum_{y} P\left(y|x\right)
\log{\frac{P\left(y|x\right)}
          {\sum_{x'} P\left(x'\right) P\left(y|x'\right)}}.
\end{equation}
Herein, $\{x\}$ is the set of symbols $m\in \{0,1,\ldots,M-1\}$,
conveyed by the $M$-ary coherent states
$\{\ket{\alpha_m}\}$,
and $\{y\}$ is the set of estimates $\hat{m}\in \{0,1,\ldots,M-1\}$. The elements of the channel matrix are given by
\beq
P(y=\hat{m}|x=m) = \bra{\alpha_x} \Pi_{y} \ket{\alpha_x},
\eeq
where
$\{\Pi_{y}\}$ is a set of detection operators.

The functional meaning of the mutual information is as follows.
Consider a block coding of length $n$ to transmit information messages
that can be represented by $M^k$ sequences of length $k$ of symbols in $\{x\}$. Here we assume $k<n$. Hence, there are $M^n$ possible sequences among whose only $M^k$ sequences are selected as codewords to represent
the information messages.
There exist $(n-k)$ redundant strings that are exploited for error correction.
The amount of information conveyed by the codewords thus constructed
is $K=k\log_2 M$ bits.
The transmission rate is then defined by
$R={K/n}=(k/n)\log_2 M$
bits/letter.
Now suppose that encoding is made under the constraint that
the frequency of $x$'s occurring in the set of codewords
is $P(x)$.
Information theory proves
\cite{Shannon48,Gallager_book,CoverThomas_book}
that by using an appropriate coding,
one can transmit the information messages
with an error probability as small as desired
if condition $R<I(X:Y)$ holds.

The capacity is defined as the maximum mutual information with
respect to the prior distribution of the letters $P(x)$
(for a memoryless channel)
\begin{equation}
C=\max_{\{P(x)\}} I(X:Y)\;.
\label{C_def}
\end{equation}

In the present context, however,
only the input variable $X$
and the corresponding set of quantum states are given.
The output variable $Y$ is to be sought for the best quantum
detection, which is described by the POVM
(positive operator-valued measure) $\{\Pi_{y}\}$.
So the capacity definition can be formulated as
\begin{equation}
C_1=\max_{\{P(x)\}}\max_{\{\Pi_y\}} I(X:Y)\;.
\label{C1_def}
\end{equation}
For the ultimate capacity, denoted $C_\infty$,
one should also consider collective decoding on
blocks of symbols.
Finding $C_1$ and $C_\infty$ for $M$-ary coherent states ($M\ge3$)
is a difficult task, and it still remains an open problem, as well as finding the maximum mutual information for a fixed $P(x)$
\begin{equation}
I_{\rm Acc} = \max_{\{\Pi_y\}} I(X:Y)\;,
\label{I_Acc}
\end{equation}
which is called the accessible information
for a given ensemble $\{\ket{\alpha_x}, P(x)\}$.

In the following we numerically evaluate the mutual information for the proposed displacement receivers and the unambiguous state discrimination
\cite{Ivanovic87Dieks88Peres88Jaeger95Chefles98,
      CheflesBarnett98,CheflesContemp}.
The former can be implemented with currently available technology, while, nowadays, the latter can be implemented in a form very close to the optimal solution \cite{van_Enk2002_USD}. In Fig.~\ref{3PSK_IXY_USD_NoF_N3_DE100DK0}
we compare, in the 3PSK case,
the mutual information attained by
the simplest 2-port scheme without feedforward,
the feedforward scheme with $N=3$ and $N=10$,
the unambiguous state discrimination (see Appendix~\ref{Appendix:USD}),
the heterodyne detection,
and the Helstrom receiver.
We observe that the USD outperforms the heterodyne limit for $\alpha^2 \ge 0.7$, but displacement receiver with feedforward is generally better.
A similar behavior is observed for the 4PSK case
reported in Fig.~\ref{4PSK_IXY_USD_NoF_N3_DE100DK0}.
These conclusions are also in agreement with the results obtained for the binary signal case \cite{Takeoka2010_cutoff_rate_JMO}.


\begin{figure}[h]
\centering
\subfigure
{
\includegraphics[width=0.95\linewidth]
{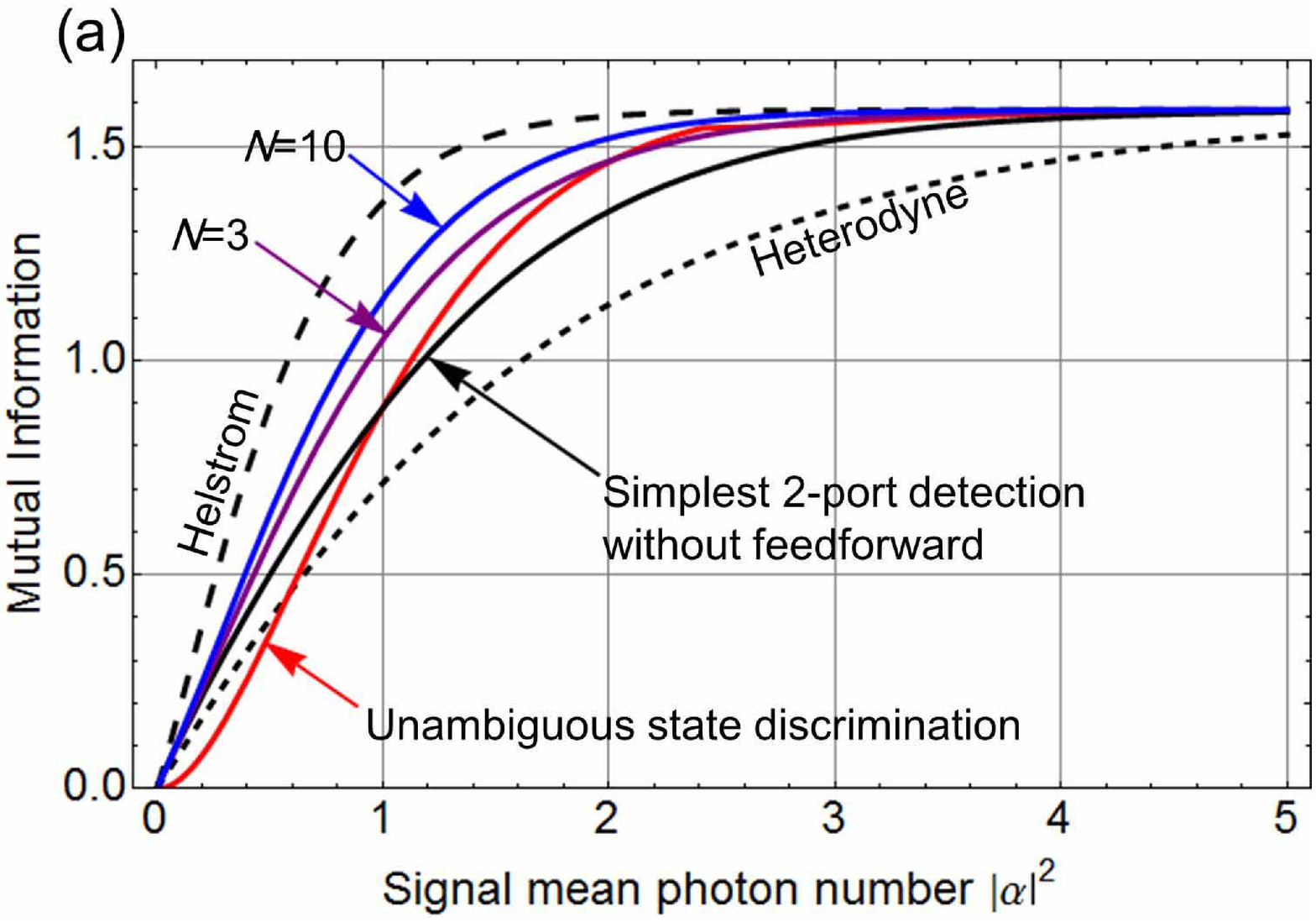}
\label{Fig3a}
}
\subfigure

{
\includegraphics[width=0.95\linewidth]
{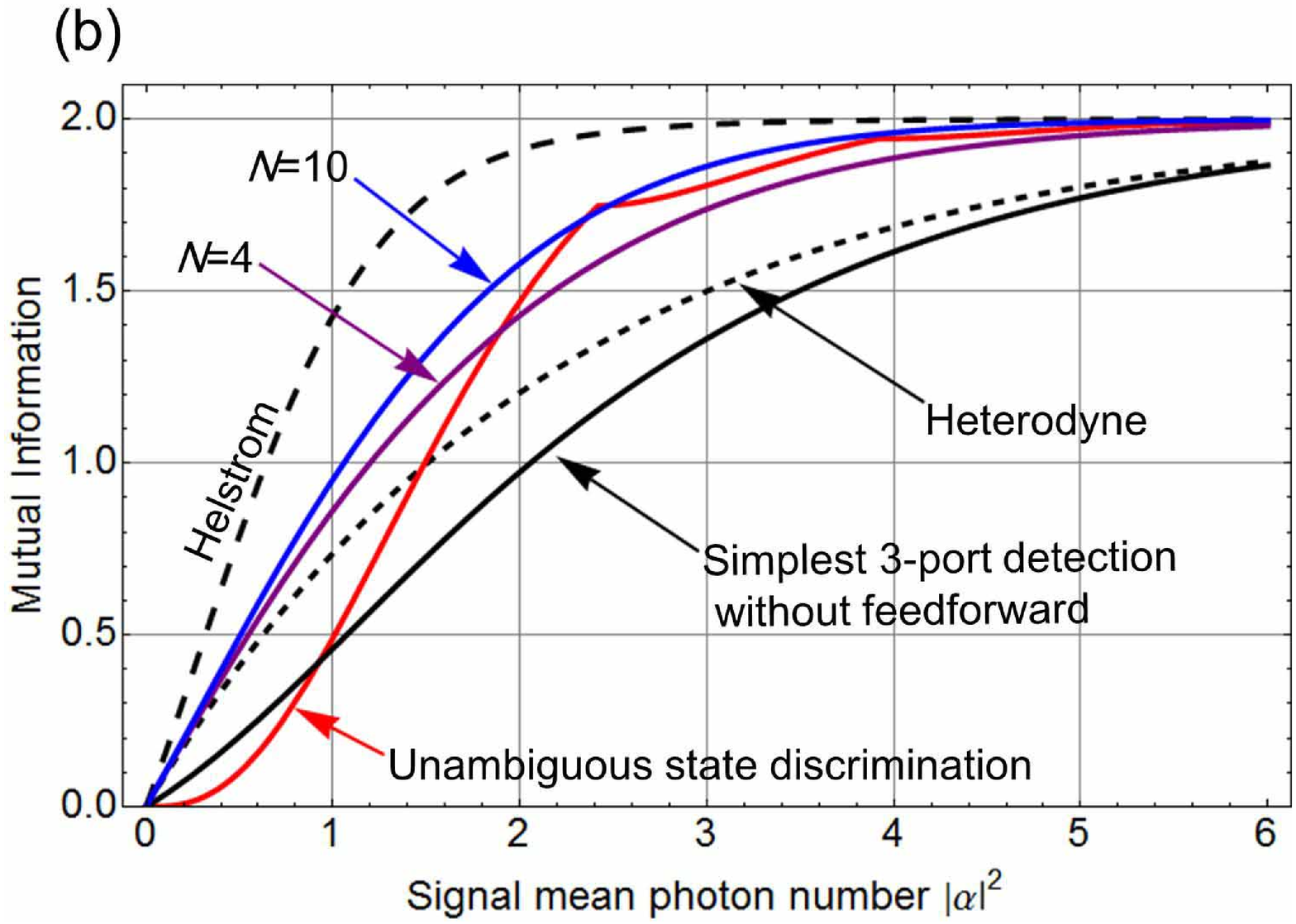}
\label{3PSK_IXY_USD_NoF_N3_DE100DK0}
}
\caption{Mutual information for (a) the 3PSK receiver 
and (b) the 4PSK receiver. 
The receivers without feedforward (black), 
the feedforward receiver 
with (a) $N=3$, 10 and (b) $N=4$, 10 
(purple and blue, respectively), 
the unambiguous state discrimination (red),
the heterodyne detection (black dotted),
and
the Helstrom receiver (black dashed).
On/off detectors are assumed to be ideal: 
$\nu=0$ and $\eta=1$.
}
\label{4PSK_IXY_USD_NoF_N3_DE100DK0}
\end{figure}


%
\section{Concluding remarks}\label{Sect:5}
%

We theoretically and numerically analyzed the performance
of the displacement receivers for the 3- and 4-PSK signals.
We showed that it could be possible to overcome the SQL,
i.e., the heterodyne limit, even without applying feedforward operations.
In particular, demonstration of the sub-SQL receiver for the 3PSK
is quite feasible with the state-of-art photon detection technologies.
We also showed that the error rate performance is drastically
increased even for moderate number of feedforward steps ($N\alt5$).
This means that the requirement for the detector
specifications can be tolerated, which would be important for the 4PSK
and agrees with the results
in \cite{Becerra_NIST_2011_MPSK_emulation_experiment}.
We also derived an asymptotic limit of the error rate with respect to $N$
and clarified the gap between our receiver and the Helstrom bound
has the order of $\alpha^2$.

The effect of feedforward also provide a remarkable gain with
respect to the mutual information in particular for $\alpha^2\le5$.
While the USD also shows a good performance
it is comparable with ($\alpha^2 \ge 2$) or lower than
($\alpha^2 \le 2$) our receiver.
Mutual information is the quantity
which eventually determines the total performance of
communication systems involving coding.
It is an important future direction to investigate the optimization
of the system with respect to mutual information,
such as the optimization of the prior probabilities or
the investigation of the better POVM consisting of
$\bar{M}$ elements with $\bar{M}>M$, as suggested by
Davies for symmetric signal sets \cite{Davies78}.

Another interesting question is whether the feedforward receiver
presented here can be applied to more general purposes
such as projecting qudit states.
For the binary case, it is known that the setup discussed 
in this paper is universal
in the sense that it can be used for arbitrary (destructive)
two-dimensional projective measurement
\cite{TakeokaSasakiLutkenhaus2006_PRL_BinaryProjMmt}.
It is a future task to generalize it to the $M$-dimensional space, 
that is, to clarify which class of the $M$-dimensional 
projection measurement can be realized 
with the present receiver setup.


\appendix

%
\section{Displacement optimization for the 3PSK feedforward receiver}
%
In the 3PSK feedforward receiver introduced in Sect.~\ref{Sect:2a}, once a photon is detected at an $(N-n)$th step, $n=1,2,\ldots,N-1$, the estimation hypothesis $\hat{\alpha}=\alpha_0$ is discharged and the estimate has to be found between symbols $\alpha_1$ and $\alpha_2$. Consequently after a photon is detected at $(N-n)$th step, a binary discrimination can be performed in the remaining $n$ steps. Hence, by using the approach in \cite{Wittmann2008_PRL_BPSK} (see also \cite{ADP11}), we fix the displacement of all the $n$ remaining steps to an optimal value $\beta^{(n)}$ that can be found by solving the following transcendent equation
$$
\frac{\sqrt{3}}{2} \sqrt{\frac{n}{N}}\,\alpha=\beta^{(n)} \tanh\left(\sqrt{3}\sqrt{\frac{n}{N}}\,\alpha \,\beta^{(n)} \right)
$$
and we decide for $\hat{\alpha}=\alpha_1$ if no photons are detected at any of the remaining $n$ steps, otherwise, if at least one photon is detected, we decide $\hat{\alpha}=\alpha_2$.

The probability of error results
\begin{eqnarray}
\tilde{P}_e & = & \frac{1}{3} e^{-3\eta\alpha^2} \left\{ 2+ \left(e^{+3\eta\alpha^2\frac{1}{N}}-1\right) \times \right.
\nonumber\\ && \left. \left[1 +
\sum_{n=1}^{N-1}  e^{+3\eta\alpha^2 \frac{n}{N}} \left( 1- \tilde{p}_1^{(n)}+\tilde{p}_2^{(n)}\right) \right]\right\}\,,\,
\label{eq:apa1}
\end{eqnarray}
where 

$
\tilde{p}_1^{(n)}=e^{-\left|\frac{\sqrt{3}}{2}\sqrt{\frac{n}{N}}\,\alpha-\beta^{(n)}\right|^2}\;,\;
\tilde{p}_2^{(n)}=e^{-\left|\frac{\sqrt{3}}{2}\sqrt{\frac{n}{N}}\,\alpha+\beta^{(n)}\right|^2}\;.
$
We note that by setting $\beta^{(n)}=\frac{\sqrt{3}}{2} \sqrt{\frac{n}{N}}\, \alpha$, i.e., by performing full symbol nulling, Eq.~\eqref{eq:apa1} becomes equal to Eq.~\eqref{eq:Pe3pskidric}.

The comparison between Eq.~\eqref{eq:apa1} and Eq.~\eqref{eq:Pe3pskidric} reveals that with this modification just a small additional gain can be obtained but only in the weak coherent state region $\alpha^2<2\,$.

%
\section{Optimization of the feedforward algorithm
via the maximization of posteriori probabilities}
\label{Appendix:opt ff}
%

Here we describe the improved feedforward algorithm used in
Fig.~\ref{4PSK_BER__BayesRule_N_step_DE100DK0}.
In Sect.\ref{Sect:3}
(except Fig.~\ref{4PSK_BER__BayesRule_N_step_DE100DK0}),
we consider the feedforward algorithm
simply change the nulling signal with the fixed ordering
conditioned on the detector click (e.g. $0\to2\to1$ for the 4PSK).
On the other hand,
the algorithm described here dynamically optimizes this ordering
with respect to the {\it posteriori} probabilities at each step.
In the following, we consider only an ideal case,
i.e. $\nu=0$ and $\eta=1$.

Suppose we start the detection process by nulling the $m=0$ signal
at the first port, detect the `off' outcome, nulling the $m=0$ signal
again at port 2, and then obtain the `on' result.
Then the input signal is guessed to be one of $m=1,2,3$ signals.
More precisely their posteriori probabilities are given as
\begin{equation}
m=0: \qquad P_0 = p_0 (1-p_0) =0 ,
\end{equation}
\begin{equation}
m=1 \, {\rm and} \, 3: \qquad P_1 = P_3 = p_1 (1-p_1) ,
\end{equation}
\begin{equation}
m=2: \qquad P_2 = p_2 (1-p_2) ,
\end{equation}
where
\begin{equation}
p_0=1 ,
\end{equation}
\begin{equation}
p_1=\mathrm{e}^{-\frac{2\alpha^2}{N} } ,
\end{equation}
\begin{equation}
p_2=\mathrm{e}^{-\frac{4\alpha^2}{N} } .
\end{equation}
These posteriori probabilities are compared to each other
and the feedforward is performed such
that the signal with the largest posteriori probability
is nulled at the next port
(if more than one signals are equally the largest,
random guess is applied).
Note that such magnitude comparison is not straightforward
as it depends on the signal power $\alpha^2$ and
the number of port $N$.

After tracing all the possible feedforward scenarios,
we find that
the success probabilities of detecting each signal are expressed as
\begin{equation}
P(0|0)=1 ,
\end{equation}
\begin{eqnarray}
&&P(1|1)=\sum_{s=1}^{N-1} d(\alpha^2,N) p_1^s(1-p_1) p_0^{N-1-s}
\nonumber
\\
&&
+\sum_{s=0}^{N -2} e(\alpha^2,N) b(N) p_1^s(1-p_1)\sum_{k=0}^{N-2-s} p_1^k (1-p_1) p_0^{N-2-s-k} ,
\nonumber
\\
\end{eqnarray}
\begin{eqnarray}
&&P(2|2)=\sum_{s=0}^{N -1} e(\alpha^2,N) a(N)p_2^s(1-p_2)p_0^{N-1-s}
\nonumber
\\
&&
+\sum_{s=1}^{\frac{N-2}{2}} d(\alpha^2,N) p_2^s(1-p_2)\sum_{k=s+1}^{N-2-s}p_1^k(1-p_1)p_0^{N-2-s-k}
\nonumber
\\
&&
+\sum_{s=1}^{\frac{N-3}{2}} d(\alpha^2,N) p_2^s(1-p_2)\sum_{k=0}^{s}p_1^k (1-p_1)
\nonumber
\\
&&
\times \sum_{l=0}^{N-3-s-k}p_1^l (1-p_1)
\nonumber
\\
&&
+\sum_{s=z(N)}^{N-3}d(\alpha^2,N)p_2^s(1-p_2)\sum_{k=0}^{N-3-s} p_1^k (1-p_1)
\nonumber
\\
&&
 \sum_{l=0}^{N-3-s-k} p_1^k (1-p_1) ,
\end{eqnarray}
\begin{eqnarray}
&&P(3|3)=
\nonumber
\\
&&
\sum_{s=0}^{N-3} e(\alpha^2,N) c(N) p_1^s(1-p_1)
\nonumber
\\
&&
\sum_{k=0}^{N-3-s} p_1^k (1-p_1)\sum_{l=0}^{N-3-s-k} p_2^l(1-p_2)
\nonumber
\\
&&
+\sum_{s=1}^{\frac{N-3}{2}} d(\alpha^2,N) p_1^s(1-p_1)
\nonumber
\\
&&
\sum_{k=s+1}^{N-3-s} p_2^k (1-p_2) \sum_{l=0}^{N-3-s-k} p_1^l (1-p_1)
\nonumber
\\
&&
+\sum_{s=1}^{\frac{N-2}{2}} d(\alpha^2,N) p_1^s(1-p_1)\sum_{k=0}^{s} p_2^k (1-p_2) p_0^{N-2-s-k}
\nonumber
\\
&&
+\sum_{s=y(N)}^{N-2}d(\alpha^2,N) p_1^s(1-p_1)\sum_{k=0}^{N-2-s} p_2^k(1-p_2) p_0^{N-2-s-k} ,
\nonumber
\\
&&
\end{eqnarray}
where $a(N),b(N),c(N)$ and $d(N)$ are
\begin{eqnarray}
a(N)=\left\{ \begin{array}{ll}
1 & s\geq (N-1) \\
0 & s<(N-1) \\
\end{array} \right. ,
\end{eqnarray}

\begin{eqnarray}
b(N)=\left\{ \begin{array}{ll}
1 & s\geq (N-2) \\
0 & s<(N-2) \\
\end{array} \right. ,
\end{eqnarray}

\begin{eqnarray}
c(N)=\left\{ \begin{array}{ll}
1 & s\geq (N-3) \\
0 & s<(N-3) \\
\end{array} \right. ,
\end{eqnarray}

\begin{eqnarray}
d(\alpha^2 , N)=\left\{ \begin{array}{ll}
1 & s\geq t(\alpha^2 ,N) \\
0 & s<t(\alpha^2,N) \\
\end{array} \right. ,
\end{eqnarray}
\begin{eqnarray}
e(\alpha^2 , N)=\left\{ \begin{array}{ll}
1 & s<t(\alpha^2,N) \\
0 &  s\geq t(\alpha^2 ,N)\\
\end{array} \right. ,
\end{eqnarray}
and
\begin{equation}
t(\alpha^2 ,N)=\frac{-2\alpha^2+N\log(1+\mathrm{e}^{\frac{2\alpha^2}{N}})}{2\alpha^2} \label{B5} .
\end{equation}
Also $z(N)$ and $y(N)$ are non-negative integers satisfying the conditions
\begin{equation}
\frac{N-3}{2}< z(N)\leq \frac{N-3}{2}+1 ,
\end{equation}
and
\begin{equation}
\frac{N-2}{2} < y(N)\leq \frac{N-2}{2}+1 .
\end{equation}
Note that we can derive such an analytical expression
only for the model without imperfections.
Because in an ideal model, the nulled signal is never be clicked which
simplify the possible feedforward scenarios and make them tractable by hand.

\vspace{0.5cm}
%
\section{Unambiguous state discrimination}
\label{Appendix:USD}
%

For completeness, we here derive the POVM for an optimal USD of
the symmetric signals.
The discussions follow \cite{CheflesBarnett98}.

In order to describe the USD we introduce a basis set,
which diagonalizes the generating operator $\hat V$
in Eq. (\ref{V_def}),
as
\beqa
\hat V&=&\exp\left( \frac{2\pi i}{M} \hat n \right),
\nonumber
\\
&=&\sum_{k=0}^{M-1} u^k \proj{\omega_k}.
\eeqa
Then one can see
\beq
\hat\rho
=\sum_{m=0}^{M-1} \proj{\alpha_m}
=\sum_{m=0}^{M-1} \lambda_m \proj{\omega_m},
\eeq
where
the eigen values $\lambda_m$ in
Eq. (\ref{Lambda_3PSK}) for the 3PSK
and
Eq. (\ref{Lambda_4PSK}) for the QPSK.

The success rate of the USD is given by
\beq
P_{USD}=\min_k{\lambda_k}.
\eeq
The detection operators are given by
\beq
\hat\Pi_m=\frac{\Lambda}{M} P_{USD}
\proj{\alpha_m^\bot}
\eeq
for the signal state $\ket{\alpha_m}$,
using the reciprocal states
\beq
\ket{\alpha_m^\bot}
=
\frac{1}{\sqrt{\Lambda}}
\sum_{k=0}^{M-1} \frac{u^{mk}}{\sqrt{\lambda_k}} \ket{\omega_k}
\eeq
where $\Lambda=\sum_k \lambda_k^{-1}$.
They satisfy the orthogonality relation
\beq
\amp{\alpha_m^\bot}{\alpha_{m'}}
=
\sqrt{\frac{M}{\Lambda}}\delta_{m,m'}.
\eeq
The operator for the inconclusive result is given by
\beq
\hat\Pi_F=\hat I -
\sum_{m=0}^{M-1} \hat\Pi_m.
\eeq
By using the POVM consisting of Eqs.~(C4) and (C7),
one can compute the mutual information for the optimal USD
system.


\end{document}